\documentclass[
    ,final            
  ,draft            
  ,numberedheadings 
  ]
  {aipproc}

\usepackage[mathscr]{eucal}
 \usepackage{amsmath, amssymb}

\def\dmf{\dot{\mathfrak{M}}}

\newcommand{\be}{\begin{equation}}
\newcommand{\ee}{\end{equation}}
\newcommand{\bdm}{\begin{displaymath}}
\newcommand{\edm}{\end{displaymath}}
\layoutstyle{6x9}

\begin{document}

\noindent Published in ~ {\it Astronomy Reports}, Volume 58, Issue 4, pp.217-227 (2014)

\vspace{1cm}

\title{A new look at Anomalous X-ray pulsars}


\author{G.S.\,Bisnovatyi-Kogan}{
  address={Space Research Institute of RAS,  84/32 Profsoyuznaya Str, Moscow 117997, Russia, and \\
National Research Nuclear University ``MEPhI'',
Kashirskoye shosse 31, Moscow 115409, Russia}
}

\author{N.R.\,Ikhsanov}{
  address={Pulkovo Observatory, Pulkovskoe Shosse 65, Saint-Petersburg 196140, Russia, and \\
  Saint Petersburg State University,  Universitetsky pr., 28, Saint Petersburg 198504, Russia}
}

\begin{abstract}
We explore a possibility to explain the phenomenon of the Anomalous X-ray Pulsars (AXP) and  Soft Gamma-ray Repeaters (SGR) within the  scenario of fall-back magnetic accretion onto a young isolated neutron star. The X-ray emission of the pulsar in this case is originated due to accretion of matter onto the surface of the neutron star from the magnetic slab surrounding its magnetosphere. The expected spin-down rate of the neutron star within this approach is close to the observed value. We show that these neutron stars are relatively young and are going through a transition from the propeller state to the accretor state. The pulsars activity in the gamma-rays is connected with their relative youth and is provided by the energy stored in the non-equilibrium layer located in the crust of low-mass neutron stars. This energy can be released due to mixing of matter in the neutron star crust with super heavy nuclei approaching its surface and getting unstable. The nuclei fission in the low-density region initiates chain reactions leading to the nuclear explosion. The outbursts are likely to be triggered by an instability developing in the region where the matter accreted by the neutron star is accumulated at the magnetic pole regions.
\end{abstract}

\maketitle


   \section{Introduction}

Anomalous X-ray Pulsars (AXPs) and Soft Gamma-ray Repeaters (SGRs) constitute a subclass
of X-ray sources with regular pulsations in their X-ray emission. One of the main
features which allows to distinguish them among other  X-ray pulsars is that all of them
are isolated objects. None of the presently known  12~AXPs and 11~SGRs is a member of a
close binary system \cite{McGill-2012}. Persistent pulsations with the period  $P_{\rm
s}$ ranging
 from 2 to 12\,seconds  and increasing with an average rate  $\dot{\nu} \sim
10^{-11}-10^{-14}\,\text{Hz\,s$^{-1}$}$, can be easily explained within the model of a
compact star with non-uniform temperature distribution over its surface and the spin
frequency  $\nu = 1/P_{\rm s}$. Relatively small value of the spin period as well as
soft X-ray spectrum and optical identification of some of these pulsars with supernova
remnants leave little doubt that AXPs and SGRs are isolated neutron stars
 \cite{van-Paradijs-etal-1995, Kaspi-2007, Mereghetti-2008,
Malov-Machabeli-2009}.

Most known so far isolated neutron stars belong to a subclass of
radio-pulsars. Similar to AXPs and SGRs, these stars are in the
spin-down state and some of them are the sources of  pulsating X-ray
emission. However, the similarity between the radio-pulsars and the
objects from the subclass of AXPs and SGRs ends here. The X-ray
luminosity of AXPs and SGRs, $L_{\rm X} \sim
10^{33}-10^{35}\,\text{erg\,s$^{-1}$}$, significantly exceeds the
spin-down rate of these neutron stars. Moreover, the unique
signature of this subclass of X-ray pulsars is recurrent gamma-ray
outbursts during which the energy of  $\sim 10^{40}- 10^{42}$\,erg
is released on the timescale of one to several seconds
\cite{Golenetskii-etal-1984, Kaspi-2004}, as well as giant gamma-ray
bursts observed from SGRs and characterized by the energy release of
$10^{43}-10^{46}\,\text{erg}$ during fractions of a second
\cite{Frederiks-etal-2007}. The pulsar luminosity during these
events significantly exceeds the Eddington  limit which eliminates a
possibility to explain gamma-ray bursts in the terms of any
accretion model. Finally, the spin periods of AXPs and SGRs range in
a narrow interval between 2 and 12 seconds that essentially
differentiates them from both ejecting and accreting pulsars whose
periods lay in the vast interval covering six orders of magnitude
(from 1.5\,ms to tens of thousands seconds).

Two most popular approaches can be distinguished among numerous
attempts of modelling AXPs and SGRs. The first one exclusively
associates the uniqueness of these objects with peculiar qualities
of the neutron star itself, while the influence of environment on
its evolution and generation of high-energy radiation is assumed to
be insignificant. The most popular within this approach is a
hypothesis about so called magnetars which considers AXPs and SGRs
as a subclass of radio-pulsars endowed with surface dipole magnetic
field as high as $10^{14} - 10^{15}$\,G. Under this assumption the
spin-down rate of a neutron star due to magneto-dipole emission is in a good
agreement with observed value, and generation of high-energy
emission from the pulsar is described in terms of decay and flaring
dissipation of its strong magnetic field \cite{Thompson-Duncan-1995,
Thompson-Duncan-1996}.

It is worth mentioning, however,  that magnetic field is not the
only possible energy source for the flaring activity of neutron
stars. As shown in \cite{Bisnovatyi-Kogan-Chechetkin-1974,
Bisnovatyi-Kogan-etal-1975, Bisnovatyi-Kogan-Chechetkin-1979}, a
non-equilibrium layer consisting of super-heavy nuclei is formed in
the interior of the neutron star during the stage of its rapid
cooling.This layer, located at the bottom of the upper crust of the
star, is in a static equilibrium at the densities
$10^{10}-10^{12}$\,g\,cm$^{-3}$. The nuclear energy stored in the
layer of the mass $\sim 10^{-4}\,M_{\odot}$, amounts to
$10^{48}-10^{49}$\,erg and can be released in nuclear explosions
caused by mixing of matter of the upper crust resulting in transfer
of super-heavy nuclei to the regions of lower density where they
become unstable. During the lifetime of the layer ($\sim 10^4$\,yr)
this amount of energy is sufficient to provide both the X-ray
luminosity of AXPs and more than 1000 recurring gamma-ray bursts
observed from SGRs. The spin-down power of the neutron star in the
initial version of this scenario (see
\cite{Bisnovatyi-Kogan-Chechetkin-1979}) was associated with the
outflow of relativistic wind generated due to nuclear explosions on
its surface.

The second approach is based on the scenario of fall-back accretion onto a young neutron
star which can occur under certain conditions soon after the supernova explosion
\cite{Colgate-1971, Zeldovich-etal-1972, Michel-1988, Woosley-Chevalier-1989,
Chevalier-1989}. In this case the star turns out to be surrounded by the residual dense
gaseous disk from which the star accretes material onto its surface
 \cite{van-Paradijs-etal-1995}. Numerical simulations presented in the papers
 \cite{Chatterjee-etal-2000, Alpar-2001}, show that over the time span of
$\sim 10^4 -10^5$\,years the rate of mass accretion from the fossil
disk onto the stellar surface remains at the level of $\dmf \sim
10^{13} - 10^{15}\,\text{g\,s$^{-1}$}$, sufficient to provide X-ray
luminosity of AXPs and SGRs. Analyzing X-ray spectra of these
sources, Tr\"umper et al.  \cite{Truemper-etal-2010,
Truemper-etal-2013} have shown that they can be well explained
within the model of disk accretion onto a neutron star with the
dipole magnetic moment $\mu = (1/2) B_* R_{\rm ns}^3$ in the range
$10^{30} - 10^{31}\,\text{G\,cm$^3$}$, where $B_*$  is the surface magnetic field of
the neutron star and $R_{\rm ns}$ is its radius. This makes it possible to
consider AXPs and SGRs as a subclass of accreting pulsars in which parameters of the
isolated neutron star are close to their canonical values.

Scenarios constructed in the frame of the second approach propose the simplest and the
most consistent description of the basic characteristics of X-ray emission from AXPs and
SGRs. At the same time they reveal their incompleteness leaving unanswered the question
about the nature of flaring activity of these objects. It is customary to avoid this
problem by making additional assumptions about very strong small-scale magnetic field on
the accreting star surface which could undergo flaring dissipation causing flaring
activity of the pulsar in gamma-rays \cite{Truemper-etal-2010, Truemper-etal-2013}.

In this paper we show that the age of such an accreting ``quasi-magnetar'' must
significantly exceed that of anomalous X-ray pulsars estimated through their spin-down
rates and the ages of associated supernova remnants. As alternative we suggest a
scenario in which flaring activity of AXPs and SGRs is explained by nuclear explosions
resulting from the release of energy stored in the non-equilibrium layer. We start
discussion of this hypothesis analyzing the structure of the fossil disk forming in the
process of fall-back accretion (see Section\,\ref{ma}). We come to the conclusion that
after the supernova explosion a neutron star turns out to be surrounded by a residual
magnetic slab of low angular momentum for a wide range of its basic parameters. In this
case,  the X-ray luminosity and spin-down rate of AXPs and SGRs can be described within
the model of accretion from the slab provided the dipole magnetic moment of the star is
$\sim 10^{29} - 10^{31}\,\text{G\,cm$^3$}$. The star with the initial (at the time of
birth)  spin period of $P_0 \geq 65$\,ms reaches this stage at a relatively young age
($10^3-10^4$\,years), beginning  its evolution in the propeller state. It transits to
the accretor state as soon as its rotation decelerates to the period in the range
$2-12$\,s. Mass accretion onto the surface of the star may be one of the main factors
contributing to the mixing of matter in its outer crust leading to release of nuclear
energy stored in the non-equilibrium layer (Section\,\ref{gamma}). The amount of this
energy is sufficient for generation of  $\sim 200$ giant gamma-rya bursts occurring at
an average  recurrence interval of about 40\,years.

Long spin periods and strong magnetic fields have been until recently believed to be the
exclusive signs of magnetars and main indicators allowing to differentiate these objects
from ordinary radio-pulsars. However, observational data  cast serious doubt on the
validity of this approach. Young et al. \cite{Young-etal-1999} reported the discovery of
a long-period radio-pulsar with the period of    8.5\,seconds. Besides, the magnetic
fields of a number of neutron stars manifesting themselves as ordinary radio-pulsars
reach values typical for hypothetical magnetars. In particular, the magnetic field
strength of the neutron star PSR J1119-6127, rotating with the period of $P_{\rm s} =
0.407$\,s, is as high as  $B = 4.1 \times 10^{13}$\,G. The magnetic field  $B = 5.5
\times 10^{13}$\,G was obtained for the radio-pulsar  PSR J1814-1744, spinning with the
period  $P_{\rm s} = 3.975$\,s \cite{Camilo-etal-2000}. Strong magnetic fields have been
detected in the radio-pulsars  (see \cite{McLaughlin-etal-2004}) PSR J1847-013 ($P_{\rm
s} = 6.7$\,s, $B = 9.4 \times 10^{13}$\,G) and PSR J1718-37 ($P_{\rm s} = 3.4$\,s, $B =
7.4 \times 10^{13}$\,G). From the other hand, the magnetic field strength of the neutron
star SGR\,0418+5729, exhibiting flaring activity which unambiguously allows to classify
this star  as an object of AXP/SGR subclass, is as weak as $B = 7.5 \times 10^{12}$\,G,
\cite{Rea-etal-2010}, which is quite typical for radio-pulsars. This means that strong
magnetic field does not constitute the necessary condition of the star's flaring
activity in the gamma-rays and cannot serve as explicit criterion for its assigning to
the subclass of AXPs/SGRs.

The main parameter determining  the difference between  SGRs and radio-pulsars within
the model of flaring activity associated  with non-equilibrium layer  is the mass of a
neutron star  \cite{bk-msu}. The mass of a non-equilibrium layer and the energy stored
there increases as the mass of the neutron star decreases so that beginning from a
certain value of the neutron star mass (unknown so far) the non-equilibrium layer starts
to exhibit flaring activity up to an appearance of  giant flares. This allows to suggest
a hypothesis that SGR-type activity characterizes low-mass neutron stars with massive
non-equilibrium layers \cite{bk-msu}. Basic conclusions of our scenario are briefly
summarized in Section\,\ref{discussion}.

 \section{Isolated accreting pulsars}\label{ma}

Modeling of supernova explosion (see, e.g.,  \cite{Moiseenko-etal-2006,
Endeve-etal-2012}) has shown that at the time of birth a neutron star is embedded in the
dense gaseous media which can essentially influence its further evolution. In
particular, under these conditions the star can switch into the ejector state only
provided the pressure of relativistic wind being ejected from its magnetosphere, $p_{\rm
rw}(r) = W_{\rm ej}/4 \pi r^2 c$, exceeds the ram pressure,  $p_{\rm ram}(r) = \rho(r)
v_{\rm ff}^2(r)$,    of the gas, located at the distance $r$, surpassing the light
cylinder radius,  $r_{\rm lc} = c/\omega_{\rm s}$.  Here $\rho$ and $v_{\rm ff} =
\left(2GM_{\rm ns}/r\right)^{1/2}$ are the density and free-fall velocity of the gas,
surrounding the neutron star of the mass  $M_{\rm ns}$. $W_{\rm ej} = f_{\rm m} \mu^2
\omega_{\rm s}^4/c^3$ is the spin-down power of the neutron star expected within a
radio-pulsar model.  $\mu$ and $\omega_{\rm s} = 2 \pi/P_{\rm s}$ are the dipole
magnetic moment and the angular velocity of the star, rotating at the period  $P_{\rm
s}$, and $f_{\rm m}$ is the dimensionless parameter which in general case is
limited as  $1 \leq f_{\rm m} \leq 4$ (see \cite{Spitkovsky-2006, Beskin-2010} and
references therein). Combining these parameters, one finds that inequality   $p_{\rm
rw}(r_{\rm lc}) > p_{\rm ram}(r_{\rm lc})$ is fulfilled if the initial spin period of
the star,  $P_0$, satisfies the condition  $P_0 < P_{\rm cr}$, where
 \be
 \label{pej}
 P_{\rm cr} \simeq 0.24\ f_{\rm m}^{2/7}\,\mu_{30}^{4/7}\,\dmf_{15}^{-2/7}\,m^{-1/7}\ \text{s}.
 \ee
Here $\mu_{30}$ and  $m$ are the initial dipole magnetic moment and
the mass of the neutron star in units   $10^{30}$\,G\,cm$^3$ and
$1.4\,{\rm M_{\odot}}$, and $\dmf_{15} =
\dmf/10^{15}\,\text{g\,s$^{-1}$}$, where $\dmf = 4 \pi r^2 \rho(r)
v_{\rm ff}(r)$ is the mass transfer rate towards the neutron star. The
surrounding gas in this case is blown away by the relativistic wind
of the star beyond its Bondi radius, $r_{\rm G} = 2 GM_{\rm
ns}/v_{\rm rel}^2$, and at early stages of its evolution the star
displays itself as a radio-pulsar. Here  $v_{\rm rel}$ is the velocity of the star in the reference frame of the
surrounding gas.

Rapid rotation is one of the necessary conditions of magnetars formation, which are
expected to have an initial spin period not exceeding a few milliseconds
\cite{Thompson-Duncan-1995}. Under these conditions, the star begins its evolution in
the ejector state for all excepted values of the parameter $\dmf$, given in the paper
 \cite{Chevalier-1989}. Its spin period in this case gradually increases according to the canonical
 radio-pulsar model and reaches the critical value at which the star switches to the propeller state
 on the time-scale
\cite{Ikhsanov-2012}
 \be
 \tau_{\rm ej} \simeq 10^6\ f_{\rm m}^{-1/2}\,I_{45}\,\mu_{30}^{-1}\,\dmf_{15}^{-1/2}\,v_8^{-1/2}\ \text{yr},
 \ee
where $I_{45} = I/10^{45}$\,g\,cm$^2$ is the moment of inertia of the neutron star and
$v_8 = v_{\rm rel}/10^8\,\text{cm\,s$^{-1}$}$. Thus, a star with initial spin period
 $P_0 < P_{\rm cr}$ can switch to the accretor state (which succeeds the propeller
 state) on the timescale comparable with the age of AXPs and SGRs (e.g.  $\le 10^4$\,yr)
only provided  $\mu \geq 10^{32}\,\text{G\,cm$^3$}$. In this light, a possibility to
form an accreting ``quasi-magnetar''  (that is a neutron star whose strong small-scale
surface magnetic field does not make significant contribution in its dipole magnetic
moment which is close to its canonical value   $\mu \sim 10^{30}\,\text{G\,cm$^3$}$,
\cite{Truemper-etal-2013}) looks rather doubtful, since the evolution time of these
stars in the ejector state exceeds  the characteristic decay time of supercritical
magnetic field estimated in \cite{Colpi-etal-2000}.

Statistical analysis of the observed radio-pulsar characteristics \cite{Narayan-1987,
Faucher-Giguere-Kaspi-2006, Rantsiou-etal-2011, Noutsos-etal-2013} shows evidence that
at the time of birth the periods of these objects are as a rule in excess of a few tens
of milliseconds.  From a theoretical point of view, this conclusion can be directly
obtained from the model of magnetorotational explosion of core-collapse supernovae
\cite{Bisnovatyi-Kogan-1970, Moiseenko-etal-2006}. The condition for a star to switch to
the ejector state in this case can be written in a form  $\dmf < \dmf_{\rm cr}$, where
 \be\label{dmfcr}
  \dmf_{\rm cr} \simeq 2 \times 10^{16}\ f_{\rm m}^{7/2}\,\mu_{30}^2\,m^{-1/2}\,\left(\frac{P_0}
  {0.1\,\text{s}}\right)^{-7/2}\,\text{g~s$^{-1}$}
  \ee
is a solution of equation $P_0 = P_{\rm cr}(\dmf)$. Otherwise, a scenario of fall-back
accretion is realized, in which the gas captured by the neutron star after its birth
forms an accretion flow approaching the star to the distance smaller than its light
cylinder radius and directly interacting with stellar magnetic field.

   \subsection{Accretion flow structure}\label{geom}

Modeling of the fall-back accretion \cite{Chevalier-1989} indicates that the matter,
captured by the neutron star after the supernova explosion, possesses a relatively low
angular momentum (its angular velocity does not exceed that of the progenitor star and
is significantly smaller than Keplerian velocity) and its accretion onto the star takes
place in a quasi-spherical regime. The duration of this phase is limited, however, by
the free-fall time of matter at the Bondi radius,
 $t_{\rm ff}(r_{\rm G}) \sim \left(r_{\rm G}^3/2 GM_{\rm
ns}\right)^{1/2}$, which for the parameters of interest  ($r_{\rm G} \leq 10^{11}$\,cm)
is negligibly small comparing to the age of AXPs. That is why, constructing the
accretion scenario for these objects is impossible without an additional assumption that
over the time span $t < t_{\rm ff}(r_{\rm G})$ the initial quasi-spherical flow is
transformed into a residual disk, in which the characteristic time of radial motion is
at least 10\,000\,years. One of indirect evidences of feasibility of this suggestion is
provided by observation of planetary system surrounding the pulsar PSR\,1257+12
\cite{Wolszczan-Frail-1992, Bisnovatyi-Kogan-1993, Wolszczan-2012}.

Modeling of a residual disk has been until recently performed in terms of Keplerian
accretion disk with initial mass
 $\sim 10^{-5}\,{\rm M_{\odot}}$ \cite{Michel-1988}.
One of possible causes of its formation is an effective turbulization  of matter,
captured by the neutron star, due to  propagation of reverse shock wave
\cite{Woosley-Chevalier-1989, Chevalier-1989}. Studies  \cite{Chatterjee-etal-2000} have
shown that such disk can donate matter to the neutron star at the rate
$10^{14}-10^{16}\,\text{g\,s$^{-1}$}$ on the timescale $10^4 -10^5$\,years, which is
sufficient to explain X-ray luminosity of AXPs and SGRs  \cite{Alpar-2001}. At the same
time, the  scenario of accretion through the Keplerian fossil disk encounters some
difficulties in interpretation of a regular spin-down of these pulsars which occurs
nowadays at a rather high rate. Such behavior of a pulsar means that the spin period of
a neutron star is significantly smaller than its equilibrium period \cite{Lipunov-1987}.
However, estimates presented in \cite{Mereghetti-Stella-1995, Alpar-2001}, suggest
otherwise. Equilibrium period of the neutron star with the dipole magnetic moment of
 $10^{30}\,\text{G\,cm$^3$}$, accreting material from the Keplerian disk at the rate
 $\sim 10^{15}\,\text{g\,s$^{-1}$}$, is close to the range of observed periods of these pulsars.
 Moreover, the Alfv\'en radius of a neutron star,
$r_{\rm A} = \left(\mu^2/\dmf \sqrt{2GM_{\rm ns}}\right)^{2/7}$, under these conditions
exceeds the corotation radius of AXPs and SGRs by a factor of 2 and turns out to be an
order of magnitude greater  than the magnetospheric radius estimated through the
temperature measurements of the black-body component in the X-ray spectra of these
sources.  These inconsistencies indicate that either X-ray emission from AXPs and SGRs
is not accretion-driven or the accretion flow structure deviates from the Keplerian
disk.

As first shown in  \cite{Bisnovatyi-Kogan-Ruzmaikin-1974,
Bisnovatyi-Kogan-Ruzmaikin-1976}, the structure of accretion flow  onto a gravitating
compact object can essentially differ from the Keplerian disk if infalling matter
possesses sufficiently strong magnetic field  $B_{\rm f}$. Rapid magnetic field
amplification  in the quasi-spherical flow leads to its deceleration and further
transformation into a magnetic disk-like envelope (magnetic slab) with low angular
momentum. This change of the flow structure occurs at the Shvartsman radius
\cite{Shvartsman-1971}
 \be\label{rsh}
 R_{\rm sh} = \beta_0^{-2/3}
 \left(\frac{c_{\rm s}(r_{\rm G})}{v_{\rm rel}}\right)^{4/3} r_{\rm G},
 \ee
at which the magnetic pressure in the accretion flow,  $E_{\rm m} = B_{\rm f}^2(r)/8\pi
\propto r^{-4}$, reaches its ram pressure, $E_{\rm ram} = \rho(r) v_{\rm ff}^2(r)
\propto r^{-5/2}$. Here $\beta_0 = E_{\rm th}(r_{\rm G})/E_{\rm m}(r_{\rm G})$, $E_{\rm
th}(r) = \rho(r) c_{\rm s}^2(r)$ is the thermal pressure and  $c_{\rm s}$ is the sound
speed in the accretion flow.

Material of the magnetic slab retains equilibrium by intrinsic magnetic field of the
accretion flow and moves towards the compact star with velocity
  $v_{\rm r} \sim r/\tau_{\rm m}$ due to dissipation of this field occurring on the
  timescale $\tau_{\rm m}$, significantly exceeding the free-fall time
  \cite{Bisnovatyi-Kogan-Ruzmaikin-1974,
Bisnovatyi-Kogan-Ruzmaikin-1976}. Basic conclusions of this scenario were confirmed by
numerical simulations of spherical accretion of magnetized plasma onto a black hole
\cite{Igumenshchev-etal-2003}.

Magnetic accretion onto a neutron star have been recently discussed in
\cite{Ikhsanov-Beskrovnaya-2012, Ikhsanov-Finger-2012, Ikhsanov-2012}. These studies
have shown that transformation of the quasi-spherical flow into a magnetic slab takes
place if the Shvartsman radius exceeds a canonical Alfv\'en radius. This condition is
valid if $v_{\rm rel} < v_{\rm ma}$, where
    \be\label{vma}
 v_{\rm ma} \simeq 1155\ \beta_0^{-1/5}\ \mu_{30}^{-6/35}\
 \dmf_{15}^{3/35}\ m^{12/35}\
 \left(\frac{c_{\rm s}(r_{\rm G})}{100\,\text{km\,s$^{-1}$}}\right)^{2/5}\ \text{km\,s$^{-1}$}.
 \ee
Interaction between the slab and the dipole magnetic field of the
neutron star results in formation of the magnetosphere, whose
minimum radius is estimated as (see Eq.~12 in
 \cite{Ikhsanov-etal-2013})
  \be\label{rma}
 r_{\rm ma} = \left(\frac{c\,m_{\rm p}^2}{16\,\sqrt{2}\,e\,k_{\rm B}}\right)^{2/13}
 \frac{\alpha^{2/13} \mu^{6/13} (GM_{\rm ns})^{1/13}}{T_0^{2/13} \dmf^{4/13}},
 \ee
where $m_{\rm p}$ and $k_{\rm B}$ are the proton mass and the Boltzmann constant, and
$T_0$ is the gas temperature at the inner radius of the slab in the region of its
interaction with magnetic field of the neutron star. A dimensionless parameter  $\alpha
= D_{\rm eff}/D_{\rm B}$, coming into this expression, determines the ratio of the
effective  coefficient of the accretion flow diffusion into the magnetic field of the
star at its magnetospheric boundary, $D_{\rm eff}$, to the Bohm diffusion coefficient
$D_{\rm B} = ckT/16eB$ \cite{Artymovich-Sagdeev-1979}, and its value, following Gosling
et al. \cite{Gosling-etal-1991}, ranges within  $0.1-0.25$. Plasma  penetrates into the
stellar magnetic field and reaches its surface in the magnetic poles regions flowing
along the field lines. The radius of the base of the accretion column arising due to
this process in a dipole magnetospheric field approximation can be estimated by an
expression $a_{\rm p} \sim R_{\rm ns} \left(R_{\rm ns}/r_{\rm ma}\right)^{1/2}$
\cite{Lipunov-1987}.

Spin evolution of the neutron star in the magnetic accretion
scenario strongly depends on the angular velocity of matter at the
inner radius of the slab. Its value in general case is limited by
inequality   $0 \leq \Omega_{\rm sl} < \omega_{\rm k}$, where
$\omega_{\rm k}(r) = \left(GM_{\rm ns}/r^3\right)^{1/2}$ is the
Keplerian angular velocity. Spindown can be observed provided
$\Omega_{\rm sl}(r_{\rm m}) < \omega_{\rm s}$
\cite{Bisnovatyi-Kogan-1991}. Absolute value of the spindown torque
applied to the neutron star from the magnetic slab under this
condition is evaluated according to expression (see Eq.~8 from
\cite{Ikhsanov-etal-2013})
  \be\label{ksds}
|K_{\rm sd}^{\rm (sl)}| = \frac{k_{\rm m}\,\mu^2}{\left(r_{\rm ma}
r_{\rm cor}\right)^{3/2}} \left(1 - \frac{\Omega_{\rm sl}(r_{\rm
m})}{\omega_{\rm s}}\right),
 \ee
where $r_{\rm cor} = \left(GM_{\rm ns}/\omega_{\rm
s}^2\right)^{1/3}$ is the corotation radius of the neutron star and
$k_{\rm m}$ is a dimensionless parameter of the order of unity.

  \subsection{Parameters of the sources}\label{m-axp}

Parameters of AXPs and the best studied SGRs are collected in Tables~\ref{axp-prop}
and \ref{sgr-prop}. We consider a situation in which these objects are isolated neutron
stars undergoing mass accretion onto their surface from the magnetic slab. The
magnetospheric radius of a neutron star in this case is denoted by expression
~(\ref{rma}). Solving inequality  $r_{\rm ma} \leq r_{\rm cor}$ for $\mu$, we find that
centrifugal barrier at the magnetospheric boundary would not prevent plasma from
reaching the stellar surface provided $\mu \leq \mu_{\rm max}$, where
 \be
 \mu_{\rm max} \simeq 10^{30}\,\text{G\,cm$^3$}\ \alpha_{0.1}^{-1}\
 T_6^{1/3}\ m^{-1/9}\ L_{34}^{2/3}\ R_6^{2/3}\ P_{\rm s}^{13/9},
 \ee
$R_6 = R_{\rm ns}/10^6$\,cm, $\alpha_{0.1} = \alpha/0.1$ and $P_{\rm
s}$ is the pulsar spin periods in seconds. This condition determines
the maximum possible value of the dipole magnetic moment of the
neutron star in the scenario under consideration. The lower limit to
the dipole magnetic moment of the neutron star undergoing spindown
at the rate $\dot{\nu}_{\rm sd}$, can be retrieved solving
inequality  $|K_{\rm sd}| \geq 2 \pi I |\dot{\nu}_{\rm sd}|$. Taking
$|K_{\rm sd}| = |K_{\rm sd}^{\rm sl}|$ (see formula~\ref{ksds}) and
solving this inequality for $\mu$, we find $\mu \geq \mu_{\rm min}$,
where
       \begin{eqnarray}\label{mumin}
  \mu_{\rm min} & \simeq & 7 \times 10^{28}\,\text{G\,cm$^3$}\ \times\ P_{\rm s}^{13/17}\ L_{34}^{6/17}\
  \dot{\nu}_{-12}^{13/17}\ \times \\
           \nonumber
   & & \times\ k_{\rm m}^{-13/17}\ \alpha_{0.1}^{9/17}\ I_{45}^{13/17}\ m^{14/17}\ T_6^{3/17}\ R_6^{2/3}.
       \end{eqnarray}

\begin{table}
\bigskip
\begin{tabular}{lcccccc}
\noalign{\smallskip}
  \hline
\noalign{\smallskip}
 {Name$^{**}$} &
 $P_{\rm s}$,\,s &
 $\dot{P},\,10^{-11}\,{\rm s/s}$ &
 $\tau,\,10^3$\,yr &
 $L_{\rm X},\,10^{34}\,\text{erg/s}$ &
 $T_{\rm bb}$,\,keV &
 $d$,\,kpc \\
\noalign{\smallskip}
  \hline
\noalign{\smallskip}
 1E\,1547.0-5408 &
 2.07 &
 4.7 &
 0.7 &
 0.08 &
 0.43 &
 4.5 \\
 CXOU\,J174505.7-381031 &
 3.83 &
 6.4 &
 0.95 &
 6.0 &
 0.38 &
 13.2 \\
 PSR\,J1622-4950 &
 4.33 &
 1.7 &
 4.0 &
 0.063 &
 0.4 &
 9 \\
 XTE\,J1810-197 &
 5.54 &
 0.8 &
 11 &
 3.9 &
 0.2 &
 3.5 \\
 1E\,1048.1-5937 &
 6.45 &
 2.3 &
 4.5 &
 0.6 &
 0.51 &
 2.7 \\
 1E\,2259+586 &
 6.98 &
 0.048 &
 230 &
 2.2 &
 0.4 &
 3.2 \\
 CXOU\,J010043.1-721134 &
 8.02 &
 1.88 &
 6.8 &
 6.1 &
 0.38 &
 60 \\
 4U\,0142+61 &
 8.69 &
 0.2 &
 70 &
 11 &
 0.4 &
 3.6 \\
 CXO\,J164710.2-455216 &
 10.61 &
 0.07 &
 230 &
 0.3 &
 0.5 &
 4 \\
 1RXS\,J170849.0-400910 &
 11.0 &
 1.9 &
 9.1 &
 5.9 &
 0.45 &
 3.8 \\
 1E\,J1841-045 &
 11.8 &
 3.93 &
 4.8 &
 19 &
 0.45 &
 8.5\\
  \noalign{\smallskip}
  \hline
     \noalign{\bigskip}
  \multicolumn{7}{l}{$^*$~~Data from electronic catalogue ``McGill SGR/AXP Online Catalog''}\\
  \multicolumn{7}{l}{http://www.physics.mcgill.ca/~pulsar/magnetar/main.html} \\
 \multicolumn{7}{l}{$^{**}$~Parameters presented in the table: $P_{\rm s}$ is the spin period of a pulsar and $\dot{P} = dP_{\rm s}/dt$ is its derivative} \\
   \multicolumn{7}{l}{$\tau = P_{\rm s}/2 \dot{P}$, $L_{\rm X}$ is the X-ray luminosity, $T_{\rm BB}$ is the temperature of blackbody component, and} \\
      \multicolumn{7}{l}{$d$ is the distance to an object}\\
\end{tabular}
\caption{Anomalous X-ray Pulsars$^*$}
\bigskip
\label{axp-prop}
\end{table}

The magnetic field strength, $B_* = 2\mu_{\rm min}/R_{\rm ns}^3$, on
the surface of AXPs and SGRs as well as the magnetospheric radius,
$r_{\rm ma}(B_*)$, and the radius of the accretion column base,
$a_{\rm p} \simeq \left[R_{\rm ns}^3/r_{\rm ma}(B_*)\right]^{1/2}$,
together with the blackbody temperature, $T_{\rm bb}$, calculated
within the magnetic accretion approach, are presented in
Table~\ref{mag}. Parameters of the objects, displayed in this Table,
favor accretion nature of their X-ray radiation. Presented values of
the magnetic field strength on the surface of these neutron stars
are in a good agreement with the observational data on their X-ray
spectra. The magnetospheric radii of these stars within our scenario
do not exceed their corotation radii, and expected spindown rates
are in accordance with the observed values. Finally, the expected
temperatures of blackbody radiation emitted at the base of the
accretion column, are also close to their observational estimates.
Somewhat larger value of this parameter estimated for a number of
sources can be connected with an effect of additional heating of the
neutron star surface by the hard radiation generated due to
comptonization of soft photons on the electrons of accretion flow.

\begin{table}
\begin{tabular}{lcccccc}
 \noalign{\smallskip}
  \hline
\noalign{\smallskip}
 {Name$^{**}$} &
 ~$P_{\rm s}$,\,s~ &
 ~$\dot{P},\,10^{-11}\,{\rm s/s}$~ &
 ~$\tau,\,10^3$\,yr~ &
 ~$L_{\rm X},\,10^{34}\,\text{erg/s}$~ &
 ~$T_{\rm bb}$,\,keV~ &
 ~$d$,\,kpc~ \\
 \noalign{\smallskip}
  \hline
 \noalign{\smallskip}
SGR\,1627-41 &
 2.6 &
 1.9 &
 2.2 &
 1.0 &
 0.5 &
 11 \\
SGR\,1900+14 &
 5.2 &
 9.2 &
 0.9 &
 9.0 &
 0.43 &
 12-15 \\
SGR\,1806-20 &
 7.6 &
 75 &
 0.16 &
 16 &
 0.6 &
 8.7 \\
SGR\,0526-66 &
 8.05 &
 3.8 &
 2.2 &
 14 &
 0.53 &
 50 \\
\noalign{\smallskip}
  \hline
     \noalign{\bigskip}
  \multicolumn{7}{l}{$^*$~~Data from electronic catalogue ``McGill SGR/AXP Online Catalog''}\\
  \multicolumn{7}{l}{http://www.physics.mcgill.ca/~pulsar/magnetar/main.html} \\
   \multicolumn{7}{l}{$^{**}$~Parameters are the same as in Table~\ref{axp-prop}} \\
\end{tabular}
\caption{Soft Gamma-Ray Repeaters$^*$}
\bigskip
\label{sgr-prop}
\end{table}

\begin{table}
\begin{tabular}{lcccccccc}
  \hline
 {Name} &
 $P_{\rm s}$, &
 $|\dot{\nu}_{\rm sd}|$, &
 $L_{\rm X},$ &
 $B_*,$  &
 $r_{\rm cor},$ &
 $r_{\rm ma},$  &
 $a_{\rm p},$   &
 $T_{\rm bb},$\\
   &
  s &
  $10^{-12}$ &
  $10^{34}$ &
 $10^{12}$\,G &
 $10^8$\,cm &
 $10^8$\,cm &
 $10^5$\,cm &
  keV   \\
   &
   &
  ${\rm Hz\,s^{-1}}$ &
  ${\rm erg\,s^{-1}}$ &
  &
  &
  &
  &
  keV   \\
 \noalign{\smallskip}
  \hline
 \noalign{\smallskip}
SGR\,1627-41 &
 2.6 &
 2.8 &
 0.25 &
 3.9 &
 3.2 &
 1.2 &
 0.9 &
 0.5 \\
SGR\,1900+14 &
 5.2 &
 3.4 &
 9.0 &
 2.7 &
 5.2 &
 1.0 &
 1.0 &
 1.1 \\
SGR\,1806-20 &
 7.6 &
 13 &
 16 &
 12 &
 6.6 &
 1.7 &
 0.8 &
 1.4 \\
SGR\,0526-66 &
 8.05 &
 0.59 &
 14 &
 1.2 &
 6.9 &
 0.6 &
 1.3 &
 1.1 \\
    \noalign{\smallskip}
  \hline
     \noalign{\bigskip}
 1E\,1547.0-5408 &
 2.07 &
 11 &
 0.08 &
 0.63 &
 2.8 &
 2.1 &
 0.7 &
 0.4 \\
J174505.7-381031 &
 3.83 &
 4.4 &
 6.0 &
 2.3 &
 4.2 &
 1.0 &
 1.0 &
 0.98 \\
J1622-4950 &
 4.33 &
 0.9 &
 0.063 &
 1.5 &
 4.6 &
 1.2 &
 0.9 &
 0.33 \\
J1810-197 &
 5.54 &
 0.26 &
 3.9 &
 0.3 &
 5.4 &
 0.46 &
 1.5 &
 0.73 \\
1E\,1048.1-5937 &
 6.45 &
 0.55 &
 0.6 &
 0.31 &
 5.95 &
 0.83 &
 1.1 &
 0.53 \\
1E\,2259+586 &
 6.98 &
 0.01 &
 2.2 &
 0.024 &
 6.3 &
 0.17 &
 2.4 &
 0.49 \\
J010043.1-721134 &
 8.02 &
 0.29 &
 6.1 &
 0.51 &
 6.9 &
 0.51 &
 1.4 &
 0.83 \\
 4U\,0142+61 &
 8.69 &
 0.03 &
 11 &
 0.11 &
 7.26 &
 0.21 &
 2.2 &
 0.77 \\
J164710.2-455216 &
 10.61 &
 0.006 &
 0.3 &
 0.011 &
 8.3 &
 0.22 &
 2.1 &
 0.32 \\
J170849.0-400910 &
 11.0 &
 0.16 &
 5.9 &
 0.4 &
 8.5 &
 0.46 &
 1.5 &
 0.81 \\
 1E\,J1841-045 &
 11.8 &
 0.28 &
 19 &
 0.99 &
 8.9 &
 0.49 &
 1.4 &
 1.1 \\
     \noalign{\smallskip}
  \hline
     \noalign{\bigskip}
   \multicolumn{9}{l}{$^*$~$|\dot{\nu}_{\rm sd}|$ is the observed spindown rate of a
pulsar,} \\
   \multicolumn{9}{l}{$B_*$ is the surface magnetic field $B_* = (1/2) \mu_{\rm min} R_{\rm ns}^3$ (see Eq.~\ref{mumin})} \\
 \multicolumn{9}{l}{$a_{\rm p}$ is the radius of the accretion column base and}\\
    \multicolumn{9}{l}{$T_{\rm bb}$ is the effective blackbody temperature $T = \left(\dfrac{L_{\rm X}}{2 \pi a_{\rm p}^2 \sigma_{_{\rm SB}}}\right)^{1/4}$} \\
\end{tabular}
\caption{Parameters of AXPs and SGRs within the model of magnetic accretion$^*$}
\label{mag}
\end{table}

  \subsection{Age and period clustering}\label{period}

As shown above, the life time of a neutron star in the ejector state under conditions of
interest significantly exceeds the age of AXPs and SGRs, evaluated from their spindown
rates and the age of associated supernova remnants. However, a much higher spindown rate
of a neutron star is expected if it starts its evolution in the propeller state. This
situation can be realized in case the initial mass accretion rate towards a neutron
star, $\dmf_0$, satisfies inequality  $\dmf_0 \geq \dmf_{\rm cr}$ (see
expression~\ref{dmfcr}). The pulsar age, $\tau$, in this case is limited solely by the
spindown time of a neutron star in the propeller state, $\tau \geq \tau_{\rm pr} = \pi
I/P_0 K_{\rm sd}^{\rm pr}$, which for the case of maximum possible value of spindown
torque, $K_{\rm sd}^{\rm pr} = \dmf\,\omega_{\rm s}\,r_{\rm m}^2$, can be  estimated as
follows:
 \be\label{taupr}
 \tau_{\rm pr} = \frac{I}{2\,\dmf\,r_{\rm m}^2}.
 \ee
Taking into account that the magnetospheric radius of a neutron star in the propeller
state is limited as   $r_{\rm m} \geq r_{\rm cor}$, we find
 \be
 \tau_{\rm pr} \simeq 1870\ I_{45}\ \left(\frac{\dmf_0}{10^{17}\,\text{g\,s$^{-1}$}}\right)^{-1}
 \left(\frac{r_{\rm cor}}{3 \times 10^8\,\text{cm}}\right)^{-2}\ \text{yr}.
 \ee
Thus, in the scenario under consideration, AXPs and SGRs are relatively young neutron
stars. Their period is likely to correspond to the period of transition of a neutron
star from propeller to accretor state \cite{Alpar-2001}. Observed period clustering in
the frame of this hypothesis indicates that the magnetospheric radius of these neutron
stars in the propeller state, $r_{\rm m}^{\rm (pr)}$, was in the range  $(2.8 - 8.9)
\times 10^8$\,cm and, hence, significantly exceeded that in the present epoch (see
Table~\ref{mag}). This conclusion does not contradict our results presented in the
previous sections. It rather points at the circumstance that the diffusion coefficient
at the magnetospheric boundary of a star in the propeller state significantly exceeds
the Bohm diffusion coefficient. This can be connected with plasma turbulization
\cite{Bisnovatyi-Kogan-Ruzmaikin-1976} by a shock wave generated at the magnetospheric
boundary of a star in the supersonic propeller state. As the diffusion coefficient
grows, the radius of stellar magnetosphere increases up to the Alfv\'en radius, $r_{\rm
A}$, whose value for the parameters of interest  ($\dmf \sim
10^{16}-10^{17}\,\text{g\,s$^{-1}$}$ and $B \sim 10^{12}-10^{13}$\,G) corresponds to the
above mentioned interval for $r_{\rm m}^{\rm (pr)}$.

\section{Flaring activity of AXPs and SGRs}
 \label{gamma}

Findings of the previous sections make it possible to explain high X-ray luminosity and rapid spindown of AXPs and SGRs within the scenario for  accretion of matter from the fossil magnetic slab  onto the surface of a young neutron star with basic parameters
close to canonical. This approach exclude a possibility to describe the pulsar activity in gamma-rays in terms of flaring dissipation of its magnetic field. The model involving non-stationary accretion turn out not to be effective because of extremely high (super-Eddington) luminosity of the sources during the  flares. The latter circumstance unambiguously indicates that the source of energy released in the course of flares is located in the star itself and, hence, is connected with peculiarities of its internal structure.

 \subsection{Energy of non-equilibrium layer}

A possibility to form a considerable reservoir of energy in the crust of a young neutron star was first pointed out in the paper by Bisnovatyi-Kogan \& Chechetkin \cite{Bisnovatyi-Kogan-Chechetkin-1974}. They have shown that a non-equilibrium layer of
super-heavy nuclei with a large energy excess relative to the equilibrium state can be generated in the envelope of a young neutron star. This non-equilibrium layer is forming at the interface between the inner and outer crust of the neutron star and at the densities  $\rho_{\rm nel} \sim 10^{10}-10^{12}$\,g\,cm$^{-3}$ remains in quasi-equilibrium state.  The initial mass of the layer is $M_{\rm nel} \sim 10^{-4}\,{\rm M_{\odot}}$ \cite{Bisnovatyi-Kogan-etal-1975}, and its life time in the absence of instabilities can be infinitely long. The total energy accumulated in the non-equilibrium layer can be as large as  $\sim 10^{49}$\,erg (see Eqs.~5.6--5.9 in \cite{Bisnovatyi-Kogan-Chechetkin-1979}).

A release of {\it free} energy stored in the non-equilibrium layer can occur as mixing of  matter in the neutron star crust proceeds and heavy nuclei from the non-equilibrium layer reach the outer layers of the star in which the density of material is  $< 10^{10}\,\text{g\,cm$^{-3}$}$. Under these conditions super-heavy nuclei become unstable to $\beta$-decay which initiate the process of nuclei fission and $\alpha$-decays, leading to chain reaction and, finally, to the nuclear explosion (see \cite{49} for detailed discussion). Duration of the energy release process in this case is determined by the characteristic time of $\beta$-decay (the slowest process in this chain), which for parameters of interest is $\tau_{\beta} \sim 10^{-4}-50$\,s (see Eq.~7.5 in \cite{Bisnovatyi-Kogan-Chechetkin-1979}). This covers the range of emission times  of gamma-ray bursts observed from SGRs, and at the same time the spectrum of primary radiation generated in nuclear explosion ranges in the energy interval $0.002-3$\,MeV. The energy efficiency of neutrons transmutation in a nucleus amounts to ~1\% of their rest energy, and $\sim 0.1\%$ of the rest energy of super-heavy nuclei can be released in the process of fission. Taking the value of efficiency parameter for the mixture of these components as  $\zeta \sim 3 \times 10^{-3}$, we find that the mass on non-equilibrium matter, necessary for appearance of gamma-burst with the energy   $\varepsilon_{\gamma}$  in the frame of nuclear explosion scenario can be evaluated as:
 \be
 M_{\gamma} \simeq 4 \times 10^{26}\ \left(\frac{\zeta}{0.003}\right)^{-1} \left(\frac{\varepsilon_{\gamma}}{10^{45}\,\text{erg}}\right)\ \text{g}.
 \ee
Thus, the energy accumulated in the non-equilibrium layer of a young neutron star proves to be sufficient for  $\sim 250$ giant ($\sim 10^{45}$\,erg) gamma-bursts, which can occur with average recurrence time of  $\sim 40$\,years over a time span of  $\sim 10^4$\,years.

  \subsection{Trigger of the flare}

\begin{figure}
 \includegraphics{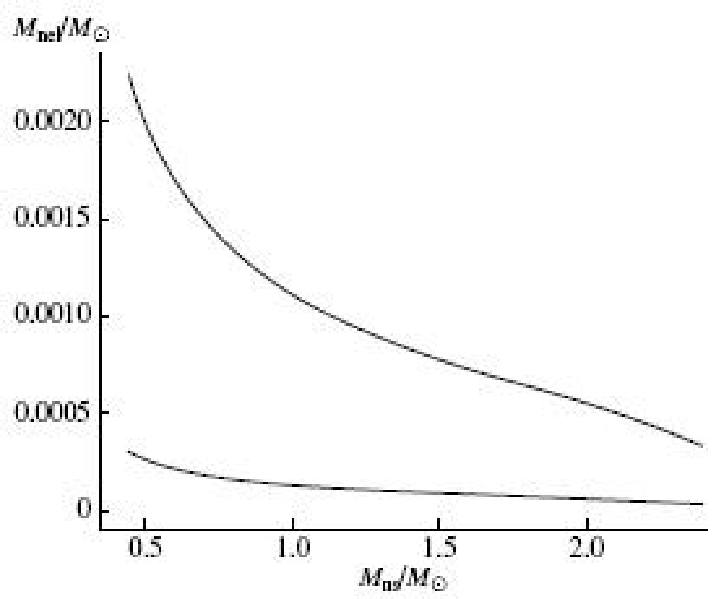}
 \caption{Dependence of mass of the non-equilibrium layer on the neutron star mass. Lines show the top and bottom boundaries of the layer mass  which is counted from the stellar surface. The state equation of equilibrium matter has been used in constructing the model  of neutron star  with boundaries of the layer setting by densities. Using the state equation accounting for non-equilibrium will increase the mass of the layer but should not essentially change the values given in the Figure.}
\label{layer-mass}
\end{figure}

According to results of calculations presented in \cite{bk-msu}, the mass of the non-equilibrium layer, forming in the crust of light neutron stars, exceeds that in more massive stars (Fig.\,\ref{layer-mass}). This allows to choose  neutron stars of moderate mass as the most probable candidates for SGRs and implies possible distinction of supernova remnants associated with the birth of these objects. The latter is in accordance with results by Marsden et al. \cite{Marsden-etal-2001} about relative compactness and higher density of supernova remnants identified with AXPs and SGRs.They have noted that observational appearance of these nebulosities may speak in favor of a relatively low energy of supernova explosion that is typical for the birth of low-mass neutron stars.

Analyzing the galactic population of neutron stars one can conclude that the youth of these objects is not a sufficient condition for realization of AXP and SGR phenomenon. Only these 23 objects among a large number of neutron stars with parameters satisfying the above mentioned criteria demonstrate flaring activity in gamma-rays. This implies that energy release from the non-equilibrium layer can neither proceed spontaneously nor be exclusively connected with peculiarities of structure evolution of the star itself (in particular, starquakes, volcanic activity and tensions arising in the stellar crust in the presence of phase transition ``liquid-gas''). The flare seems to be triggered by the action of external factors on the star, one of which may be accretion of matter onto its surface.

\subsection{Possible model of a low-mass neutron star formation}

The birth of a low-mass neutron star should be a rather rare event, since a fraction of AXP/SGRs among currently known neutron stars is as low as one percent (see, e.g.  \cite{McGill-2012,rp}). Detection of binary pulsars in systems with neutron stars enabled accurate determination of their masses. For instance, the observations  of a binary pulsar system J 1518+4904 revealed with probability 95.4\% that the masses of components are $m_{\rm p} = 0.72^{+0.51}_{-0.58} M_\odot$ and $m_{\rm e} = 2.00^{+0.58}_{-0.51} M_\odot$ \cite{nsmass1}.

The most accurate measurements of neutron stars masses were carried out for two close binaries consisting of
neutron stars, in which effects of general relativity are the most clearly displayed.
In one of them, containing the first binary pulsar discovered, the mass of the visible recycled pulsar is
1.4424\,$M_\odot$, while the mass of its invisible companion is 1.386\,$M_\odot$ \cite{dp1}.
In another double pulsar system the mass of the recycled pulsar is
 1.3381\,$M_\odot$, and the mass of the young ``normal'' pulsar is
 1.2489\,$M_\odot$ \cite{dp2}.
 As we can see from above, the mass of the recycled pulsar in both systems exceeds the mass of its companion which
 has not undergone mass accretion. In the process of disk accretion the recycled pulsar
 mass is increased by an amount which does not  depend on the value of its initial period (exceeding $\sim 0.2$\,s)
 and is related to the final spin period of a pulsar $P_{ms}$ (in milliseconds) as \cite{tauris12}

\be\label{rpmass}
 \Delta m =0.22\,M_\odot\ \frac{(M/M_\odot)^{1/3}}{P_{ms}^{4/3}}.
 \ee

A pulsar with a final mass of $1.4\,M_\odot$ and a recycled spin period of either 2\,ms, 5\,ms, 10\,ms or 50\,ms requires to accrete an amount of mass 0.1, 0.03, 0.01 and 0.001\,$M_\odot$, respectively. Thus, in double neutron stars systems with spin
periods in excess of 23\,ms (see, e.g. \cite{bk06}), the amount of mass added to an accreting pulsar during recycling  is
insignificant.

Neutron stars are born in core-collapse supernova explosion. Collapse results from the loss of hydrodynamic stability at final stage of evolution of massive stars. The majority of neutron stars observed as radio-pulsars and X-ray sources are likely to be born in this process. We assume that single low-mass neutron stars, which are suggested to be SGRs, are formed through another, rarely realized channel. As example let us consider the evolution of a single star of moderate mass, which ends with loss of thermal stability of a star and off-center explosion of its degenerate core. The models  with off-center explosion and collapsing core were discussed in \cite{bn86} for supernovae of Type\,Ib. An outcome of this process must depend on the mass ratio between the iron core and the shell with He and C--O layers. The mass of collapsing remnant should be close to that of an iron core which can amount to $0.4\,-\,0.6\,M_\odot$. When the whole star (core + shell) looses its hydrodynamic stability, the collapse starts, the temperature in He and C--O  shells exceeds the critical value that gives rise to  thermonuclear explosion leading to ejection of both shells. As a result of this process a low-mass neutron star can be born. In is necessary to note, however, that a definite answer to the question about a possibility to realize this scenario requires  numerical simulations of stellar evolution and, in particular, of the explosion dynamics.

   \section{Conclusions}\label{discussion}

The basic conclusion of this paper is that it is not necessary to invoke the magnetar hypothesis
in order to explain the spin evolution of AXPs and SGRs. Expected spindown rate of a neutron star accreting matter from a non-Keplerian magnetic slab corresponds to the observed value provided the magnetic field strength on its surface is in the range
 $10^{10} - 10^{13}$\,G. The blackbody temperature of the pulsar X-ray radiation evaluated within this approach  is also close to the observed estimate that speaks in favor of  feasibility of the proposed scenario.

Flaring activity of a neutron star in gamma-rays in the absence of super-strong magnetic field can be explained assuming an existence of energy reservoir located under its surface. As such we consider a non-equilibrium layer of super-heavy nuclei which is forming at early stages of a neutron star evolution. In the frame of this approach, AXPs and SGRs turn out to be relatively light neutron stars undergoing accretion of matter onto their surface which lead to development of instabilities in their crust and
can  trigger  their flaring activity.

Formation of a low-mass neutron star could be connected with off-center nuclear explosion and induced collapse of the star
central region in supernova explosion. Discovery of neutron stars with masses close to one solar mass evidences that the mass of the remnant formed in the process of collapse and supernova explosion can be significantly smaller than the initial mass of the collapsing core. This is an indirect confirmation of possibility to form low-mass neutron stars.

\begin{theacknowledgments}
The authors are grateful to S.O.\,Tarasov and N.G.\,Beskrovnaya  for their help in this work.
NRI acknowledges the support of the RAS  Presidium Program  N\,21 ``Non-stationary phenomena in the Universe'',
RFBR under grant N\,13-02-00077 and President Support Program for Leading Scientific Schools  NSH-1625.2012.2. GSBK acknowledges the support of RFBR under grant 11- 02-00602, the RAS Program  ``Formation and evolution of stars and galaxies'' and President Support Program for Leading Scientific Schools  NSH-5440.2012.2.
\end{theacknowledgments}

\end{document}